\documentclass[fleqn,10pt]{wlscirep}
\usepackage[utf8]{inputenc}
\usepackage[T1]{fontenc}
\usepackage[right]{lineno}
\usepackage{booktabs}
\usepackage{multirow}
\usepackage{makecell}

\usepackage[nottoc]{tocbibind}

\newcommand{\lp}[1]{\left( #1 \right)}
\newcommand{\lb}[1]{\left[ #1 \right]}

\newcommand{\Rs}[1]{R_S\left( #1 \right)}

\newcommand{\hast}{h^{\ast}}
\newcommand{\chiC}[1]{\chi_C\left( #1 \right)}

\newcommand{\Rtilde}[1]{\widetilde{R}\left( #1 \right)}

\title{Incomplete Reputation Information and Punishment in Indirect Reciprocity}

\author[1,$\dagger$]{Heejeong Kim}
\author[2,3,*]{Yohsuke Murase}

\affil[1]{Department of Physics, Pukyong National University, Busan, 48513,  Korea}
\affil[2]{RIKEN Center for Interdisciplinary Theoretical and Mathematical Science (iTHEMS), Wako, 351-0198, Japan}
\affil[3]{RIKEN Center for Computational Science, Kobe, 650-0047, Japan}

\affil[$\dagger$]{kimhj9507@gmail.com}
\affil[*]{Corresponding author: yohsuke.murase@riken.jp}

\begin{abstract}
Indirect reciprocity promotes cooperation by allowing individuals to help others based on reputation rather than direct reciprocation.
Because it relies on accurate reputation information, its effectiveness can be undermined by information gaps.
We examine two forms of incomplete information: incomplete observation, in which donor actions are observed only probabilistically, and reputation fading, in which recipient reputations are sometimes classified as ``Unknown''.
Using analytical frameworks for public assessment, we show that these seemingly similar models yield qualitatively different outcomes.
Under incomplete observation, the conditions for cooperation are unchanged, because less frequent updates are exactly offset by higher reputational stakes.
In contrast, reputation fading hinders cooperation, requiring higher benefit-to-cost ratios as the identification probability decreases.
We then evaluate costly punishment as a third action alongside cooperation and defection.
Norms incorporating punishment can sustain cooperation across broader parameter ranges without reducing efficiency in the reputation fading model.
This contrasts with previous work, which found punishment ineffective under a different type of information limitation, and highlights the importance of distinguishing between types of information constraints.
Finally, we review past studies to identify when punishment is effective and when it is not in indirect reciprocity.
\end{abstract}

\begin{document}

\flushbottom
\maketitle
\thispagestyle{empty}


\section*{Introduction}\label{sec:intro}
\noindent

Maintaining cooperation is a fundamental challenge in animal and human societies, as cooperation is often vulnerable to exploitation by selfish individuals~\cite{rand:TCS:2013,melis:ptrs:2010}.
Indirect reciprocity is a key mechanism for sustaining cooperation, where individuals help others based on their reputation rather than direct reciprocation~\cite{nowak2005evolution,wedekind2002long,sigmund2012moral,santos2021complexity,okada2020review}.
In this framework, those who engage in altruistic behaviors obtain a good reputation and may receive future rewards from third parties because of that good reputation.
In such a society, cooperative behaviors are rewarded in the long term, thereby stabilizing cooperation.

In order to promote cooperation in indirect reciprocity, individuals often rely on social norms that dictate how they should behave based on the reputations of others and how they should assess the actions of others.
The rule that determines the action taken by a donor based on the recipient's reputation is called the action rule, while the rule that determines how the donor's action is assessed is called the assessment rule.
The combination of these two rules is referred to as a social norm.
Only under appropriate social norms can cooperation be evolutionarily stable, and the central question is which social norms can maintain cooperation.
In early studies of indirect reciprocity, the action and assessment rules were assumed to be first-order, meaning that the action rule depends only on the recipient's reputation and the assessment rule depends only on the recipient's reputation and the donor's action~\cite{nowak1998evolution,nowak2005evolution}.
While these first-order social norms, such as Image Scoring, are simple, they are not sufficient to maintain cooperation, demonstrating that more complex social norms are needed~\cite{leimar2001evolution,panchanathan2003tale}.
To stabilize cooperation, the context of the donor's action must be considered, which led to the introduction of second-order and third-order social norms~\cite{ohtsuki2004should}.
In a seminal paper by Ohtsuki and Iwasa~\cite{ohtsuki2004should}, third-order social norms were comprehensively enumerated and analyzed, leading to the discovery of the ``leading eight'' social norms.
Subsequently, numerous studies have been conducted based on these findings, exploring the dynamics of indirect reciprocity under various conditions and assumptions~\cite{ohtsuki2006leading,ohtsuki2007global,ohtsuki2009indirect,pacheco2006stern,santos2016social,santos2018social,masuda2012ingroup,berger2011learning,clark2020indirect,hilbe2018indirect,uchida2010effect,kawakatsu2024mechanistic,kessinger2023evolution,fujimoto2024leader,schmid2023quantitative,michel2024evolution,murase2024computational,murase2024indirect}.
More recently, this model has been extended to include stochastic reputation updates, with the classical deterministic model as a special case~\cite{murase2023indirect,schmid2021unified,murase2025costly,glynatsi2025exact}.
This extension facilitates analytical calculations and enables a more comprehensive analysis.
The necessary and sufficient conditions for the evolution of cooperation under these extended models have been derived~\cite{murase2023indirect,glynatsi2025exact}.
This analysis can be applied to study models with additional actions, such as costly punishment~\cite{murase2025costly}.

An important aspect of indirect reciprocity is the role of information limitations in shaping social norms and cooperation dynamics~\cite{nowak2005evolution,masuda2007tag,nakamura2011indirect,suzuki2013indirect,yoeli2013powering,hilbe2018indirect,michel2024evolution}.
Since it relies on reputation information to sustain cooperation, the effectiveness of indirect reciprocity is contingent on the availability and reliability of this information.
In small societies where everyone knows each other, reputation information is more likely to be accurate and up-to-date.
However, as societies grow larger and interactions become more anonymous, actions are not always witnessed, and reputation information may be inherently limited and uncertain.
Maintaining cooperation in such environments is a significant challenge, as it becomes difficult to identify free-riders and enforce social norms.

\begin{figure}[!h]
\centering
\includegraphics[width=0.9\textwidth]{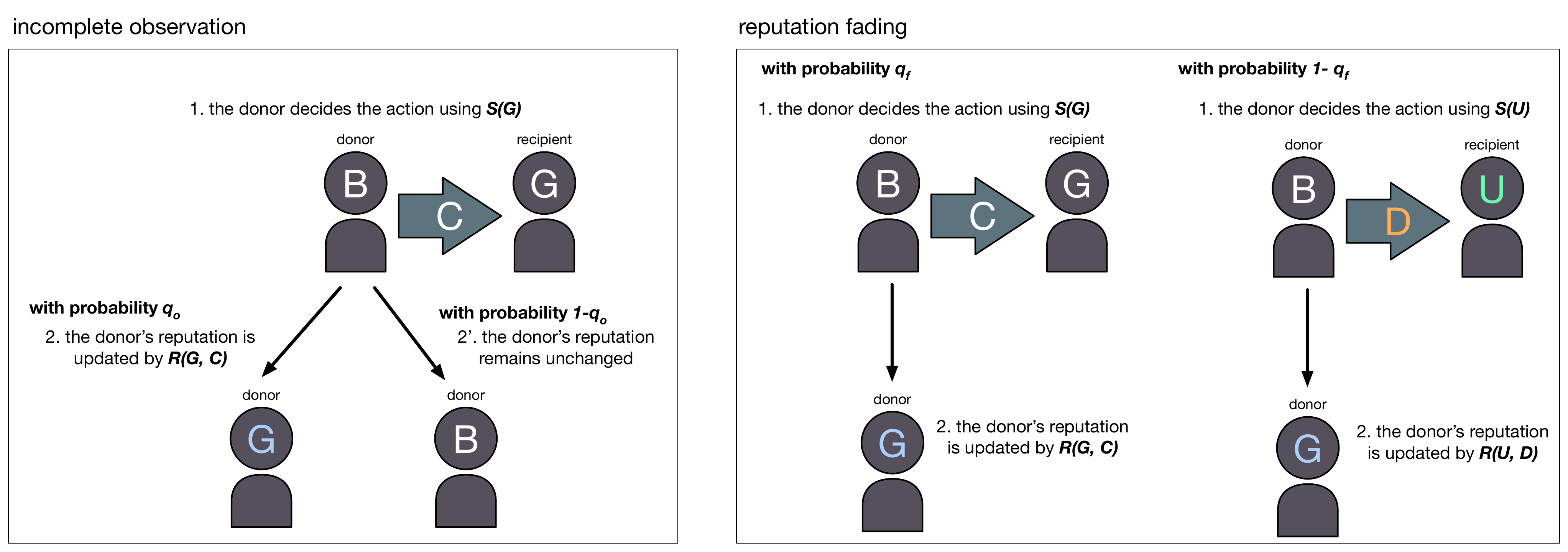}
\caption{
  A schematic diagram of the models considered in this paper.
  Suppose that a $B$ donor meets a $G$ recipient.
  (Left) In the incomplete observation model, the donor's action is not always observed.
  With probability $q_o$, the donor's action ($C$) is observed and the donor is assigned a new reputation $G$ according to $R(G, C)$.
  With probability $1-q_o$, the action is not observed and the donor's reputation remains unchanged despite the donor's cooperation.
  (Right) In the reputation fading model, the recipient's reputation is not always accessible.
  With probability $q_f$, the recipient's reputation is correctly identified as $G$.
  The donor decides to cooperate following her action rule $S(G)$, and the donor is assigned a new reputation $G$ according to $R(G, C)$.
  With probability $1-q_f$, the recipient's reputation is not accessible.
  The donor decides her action based on $S(U)$, and she is assigned a new reputation $G$ according to $R(U, D)$.
}
\label{fig:overview}
\end{figure}

In this paper, we compare two models that capture different aspects of information limitations in indirect reciprocity within the public assessment model.
See Figure~\ref{fig:overview} for a schematic overview of the models.
In the first model, we consider \textit{incomplete observation}, where donor actions are observed only with a certain probability.
When the donor's action is not observed, the donor's reputation remains unchanged, leading to sporadic reputation updates.
Specifically, the reputations are not always up to date, and the donor's action is not always assessed.
In the second model, we consider \textit{reputation fading}, where the recipient's reputation is not always accessible~\cite{nowak1998evolution,nakamura2011indirect}.
With probability $q_f$, the recipient's reputation is identified as either good ($G$) or bad ($B$), while with probability $1-q_f$, the recipient's reputation is considered unknown ($U$).
The unknown reputation state is explicitly introduced, requiring the social norm to incorporate it into both action and assessment rules.
While these models have been studied in the literature, we revisit them using recently proposed analytical frameworks for indirect reciprocity~\cite{glynatsi2025exact,murase2023indirect,murase2025costly}.
Although the distinction between these two models may seem subtle, we show that they lead to qualitatively different conditions for cooperation stability.

Our analysis reveals that reputation fading creates severe barriers to cooperation maintenance, as it becomes harder to identify free-riders.
To address this challenge, we investigate the role of costly punishment as a third action, Punishment ($P$), alongside Cooperation ($C$) and Defection ($D$)~\cite{ohtsuki2009indirect,murase2025costly}.
Here, $P$ is modeled as an action that incurs a cost $\alpha$ for the donor while reducing the recipient's payoff by $\beta$.

The effectiveness of costly punishment in human cooperation has been debated in the literature~\cite{boyd1992punishment,henrich2001people,boyd2003evolution,brandt2006punishing,mathew2011punishment,guala2012reciprocity,raihani2015reputation,raihani2015third,raihani2019punishment}.
On one hand, punishment increases the cost of maintaining a bad reputation, potentially deterring free-riding behavior.
However, because punishment is costly for the donor, it creates a second-order free-riding problem where individuals may avoid the expense of enforcing social norms.
Furthermore, previous work has shown that only limited parameter regions exist where costly punishment is effective in indirect reciprocity~\cite{ohtsuki2009indirect}.
In environments with inaccurate reputation information, punishments are carried out more frequently, and the overall efficiency often becomes worse than that under Always Defect (ALLD).
As a result, there are limited parameter regions where costly punishment is effective.
Conversely, recent studies have demonstrated that punishment becomes particularly effective when defection is difficult to detect with certainty~\cite{murase2025costly}.
Punishment may or may not be effective in indirect reciprocity, depending on the model and parameter settings, and the question of when costly punishment is effective remains an open question.

Building on this foundation, we demonstrate that costly punishment is especially valuable under reputation fading.
Our key finding is that while traditional norms without $P$ require higher benefit-to-cost ratios as identification probability decreases, norms with $P$ can maintain cooperation across broader parameter ranges.
This effectiveness stems from punishment's ability to create stronger incentives for reputation management when reputation information is uncertain.
In the Discussion section, drawing on past studies, we provide a more general discussion of when costly punishment is effective and when it is not in indirect reciprocity.

\section*{Models}\label{sec:model}

\subsection*{Baseline model}

In this study, we develop models based on the basic framework of Ohtsuki and Iwasa~\cite{ohtsuki2004should}.
In this baseline model, we consider an infinitely large population of players who interact in pairwise donation games.
In each round, two players are randomly chosen, one as a donor and the other as a recipient.
The donor decides whether to cooperate ($C$) or to defect ($D$) based on the recipient's reputation.
Cooperation incurs a cost $c > 0$ for the donor and results in a benefit $b>c$ for the recipient.
Defection leads to a payoff of zero for both players.
If the donation game is only played once, the donor is better off by defecting, creating a social dilemma.
However, we assume that population members play many donation games against different opponents.
In this case, individuals can build up a reputation over time, which may influence their cooperative behavior.

In this baseline model, we assume that reputations are binary and public.
Specifically, the reputation of a player can be either good ($G$) or bad ($B$), and it is known to all other players without any disagreement.
How players form reputations, and how they act based on these reputations, depends on their social norm.
Throughout this paper, we consider second-order social norms, where the assessment rule depends on the recipient's reputation and the donor's action but not on the donor's reputation.
An action rule $S(X)$ determines the action taken by a donor based on the recipient's reputation $X \in \{ G, B \}$.
It returns either $C$ or $D$.
An assessment rule $R(X, A)$ is a function that returns the probability that a donor is assessed as $G$ after taking action $A \in \{ C, D \}$ against a recipient with reputation $X$.
We consider stochastic assessment rules, where $R(X, A)$ can take any value in the range $[0, 1]$, which includes deterministic rules as a special case.
We assume without loss of generality that the action rule is deterministic while the assessment rule is stochastic, since non-deterministic action rules are generally not evolutionarily stable~\cite{murase2023indirect}.

The assessments are subject to errors that occur when new reputations are assigned.
With probability $\mu$, the assignments are the opposite of those prescribed by the social norm.
As a result, instead of their intended assessment rules, players implement the following effective assessment rule
\begin{equation}
  \widetilde{R}\lp{X,A} = \lp{1-\mu}R\lp{X,A} + \mu\lb{1-R\lp{X,A}}.
\end{equation}
With these errors, the assessment rule is constrained as $\mu \leq \widetilde{R}(X,A) \leq 1 - \mu$.
For theoretical analysis, it is important that $\mu > 0$ to ensure ergodicity of the model dynamics.
When $\mu > 0$, the population's reputation distribution converges to a unique stationary state irrespective of the initial state~\cite{murase2023indirect}.
We consider the limit where $\mu$ is sufficiently small but positive ($\mu \to 0$) in the following analysis, which is a common assumption in the literature on indirect reciprocity~\cite{ohtsuki2004should,ohtsuki2006leading,murase2023indirect}.
Hereafter, $\widetilde{R}(X,A)$ is simply denoted as $R(X,A)$, since $\widetilde{R}(X,A) \to R(X, A)$ as $\mu \to 0$.

Based on this baseline model, we develop extensions that capture increasingly realistic departures from perfect information.
Model 1 examines the effect of incomplete observation while maintaining perfect reputation access.
Model 2 introduces unknown $U$ reputation states and a costly punishment action $P$.
This progression allows us to isolate the effects of different types of information limitations.

\subsection*{Model 1: Incomplete observation}

The first model we consider is a naive generalization of the baseline model to the case of incomplete observation.
In this model, the donor's action is not always observed.
The donor's action is observed and their reputation is updated only with probability $q_o$.
Otherwise (with probability $1-q_o$), the donor's reputation remains unchanged.
See the left panel of Fig.~\ref{fig:overview}.

This model has been considered in several previous works.
We revisit this model using the latest theoretical understanding of indirect reciprocity and show that introducing incomplete observation leads to no significant changes in the results, which may seem counterintuitive at first.
As we demonstrate in Section \textit{Results}, this model yields identical ESS conditions and cooperation levels to the baseline model.

\subsection*{Model 2: Reputation fading}

Model 1 assumes that reputations remain unchanged even when actions are not observed.
A more realistic assumption would be to disregard reputation and treat players who are not observed for a while as unknown.
As the second model, we consider a model introduced by Nakamura and Masuda~\cite{nakamura2011indirect}.
When the donor interacts with the recipient, the recipient's reputation is not always available.
With probability $q_f$, the recipient's reputation is correctly identified as either $G$ or $B$ by the donor and the observers who update the donor's reputation.
With probability $1-q_f$, the recipient's reputation is not available.
Instead of $G$ or $B$, the donor and the observers consider the recipient's reputation as Unknown ($U$).
Note that the accessibility of the recipient's reputation is synchronized between the donor and the observers.
When the recipient's reputation is accessible, both the donor and the observers identify the recipient's reputation; otherwise, both recognize the recipient's reputation as $U$.
While this assumption is referred to as ``concomitant observation'' in previous literature~\cite{nakamura2011indirect}, we call it ``reputation fading'' in this paper to avoid confusion with the incomplete observation model defined above.

Since a new reputational state $U$ is introduced, both assessment and action rules are extended for $U$ reputation.
The donor's action rule $S(X)$ defines the action taken by the donor when the recipient's reputation is $G$, $B$, or $U$.
The same holds for the assessment rule $R(X, A)$.

This model has been analyzed in the previous literature~\cite{nakamura2011indirect}.
The analysis showed that reputation fading makes cooperation harder to maintain, requiring larger benefit-to-cost ratios $b/c$ to maintain cooperative ESS norms.
In our analysis, another action $P$ is introduced to investigate the role of costly punishment in this model.
When the donor chooses $P$, the recipient's payoff is reduced by $\beta > 0$ and the donor incurs a cost $\alpha > 0$.
The action rule is thus extended to $S(X): \{G, B, U\} \to \{C, D, P\}$, and the assessment rule becomes $R(X, A): \{G, B, U\} \times \{C, D, P\} \to [0,1]$.
This extension allows us to investigate whether costly punishment can mitigate the cooperation challenges introduced by reputation fading.

While this model assumes that the recipient's reputation is unknown with probability $1-q_f$, this is equivalent to considering that the donor's reputation is assigned $U$ with probability $1-q_f$ after the interaction.
Since we consider second-order social norms, a new reputation assigned to the donor is completely independent of the donor's previous reputation.
In other words, the assigned reputation lasts on average only for one round as a recipient.
Therefore, we can equivalently consider that the player's reputation is $U$ when chosen as a recipient, or that it was assigned $U$ in the previous round as a donor.
These two interpretations yield mathematically the same results in our analysis.

\section*{Results}\label{sec:result}

\subsection*{Incomplete observation model}

Our first key finding is that incomplete observation ($q_o < 1$) has no effect on either cooperation levels or the conditions for the evolutionary stability of social norms.
This surprising result holds for both second- and third-order social norms.
We demonstrate this using analytical results from recent advances in indirect reciprocity theory~\cite{glynatsi2025exact}.
Below, we provide an intuitive explanation for second-order norms, with the formal third-order analysis presented in Appendix.

To build intuition, consider the standard discriminator action rule ($S(G) = C, S(B) = D$) under complete observation ($q_o = 1$).
For any social norm to be evolutionarily stable, the action rule must prescribe the optimal response in every possible context.
According to~\cite{glynatsi2025exact}, this occurs if and only if
\begin{equation}
  \label{eq:ESS_2nd}
  \begin{split}
    \lb{R\lp{G,C} - R\lp{G,D}} \Delta v > c \\
    \lb{R\lp{B,C} - R\lp{B,D}} \Delta v < c
  \end{split}
\end{equation}
where $\Delta v$ represents the long-term payoff advantage of having a good versus bad reputation.
For second-order norms, $\Delta v = b$ since reputations last only one round as recipients: $G$-players receive benefit $b$ when chosen as recipients, while $B$-players receive nothing.

The intuition behind Eq.~\eqref{eq:ESS_2nd} is straightforward: the left-hand side represents the expected long-term reputational benefit from cooperation versus defection, while the right-hand side represents the immediate cost $c$.
Cooperation is optimal when the long-term reputational benefits outweigh immediate costs; otherwise, defection is preferred.
When the recipient is $G$, the left-hand side must be greater than the right-hand side for cooperation to be favored.
On the other hand, when the recipient is $B$, the left-hand side must be less than the right-hand side for defection to be favored.

Now consider incomplete observation where actions are observed with probability $q_o < 1$.
Two competing effects emerge:
(1) Reputations persist longer (expected duration $1/q_o$ instead of $1$), increasing the value of a good reputation by a factor of $1/q_o$;
(2) The probability of reputational assessment decreases by a factor of $q_o$, thereby reducing the expected impact.

These effects exactly cancel out: while the stakes of each assessment increase ($\Delta v \to \Delta v/q_o$), the frequency of assessment decreases (by a factor of $q_o$), leaving the overall ESS conditions unchanged.
Therefore, incomplete observation has no effect on cooperation or evolutionary stability.

This cancellation result extends to third-order social norms, where assessment rules depend on both donor and recipient reputations.
The same mathematical mechanism applies: reduced observation frequency is exactly compensated for by increased reputational stakes.
See Appendix for a formal proof for general third-order norms.

\subsection*{Reputation fading model}

In contrast to incomplete observation, reputation fading fundamentally alters cooperation dynamics.
We begin by analyzing how reputation distributions evolve in this model.

Let $h$ denote the fraction of individuals with a good ($G$) reputation in a homogeneous population.
The reputation dynamics are governed by:
\begin{equation}
    \dot{h} = q_f \lb{ h \Rs{G} + \lp{1-h} \Rs{B} } + \lp{1-q_f} \Rs{U} - h,
\end{equation}
where $\Rs{X} \equiv R(X, S(X))$ is the probability that a donor receives good reputation when interacting with a recipient of reputation $X$.

This equation captures two scenarios: with probability $q_f$, reputation is identified correctly and interactions occur with $G$ or $B$ recipients (fractions $h$ and $1-h$, respectively); with probability $1-q_f$, reputation is unknown and all interactions are treated as involving $U$ recipients.

At equilibrium ($\dot{h} = 0$), the steady-state fraction of individuals with good reputations is:
\begin{align}\label{eq:hast}
    \hast = \frac{\lp{1-q_f} \Rs{U} + q_f\Rs{B}}{1 - q_f\Rs{G} + q_f\Rs{B}}.
\end{align}
This formula applies for $q_f < 1$; the case $q_f = 1$ reduces to standard models analyzed elsewhere~\cite{glynatsi2025exact,murase2024indirect,ohtsuki2004should}.

Using $\hast$, the cooperation probability of the population is given by
\begin{equation}\label{eq:pc_synchronous}
    p_c = q_f \lb{ \hast \chiC{G} + \lp{1-\hast} \chiC{B} } + \lp{1-q_f} \chiC{U},
\end{equation}
where $\chiC{X}$ is the indicator function that takes the value 1 if the action rule $S$ prescribes cooperation with reputation $X \in \{ G, B, U \}$, and 0 otherwise.

We are interested in norms that maintain evolutionarily stable cooperation.
We require a social norm to have $p_c = 1$ in the steady state and to be evolutionarily stable against mutations with different action rules.
We call such norms cooperative ESS (CESS) norms.

For full cooperation, $\chiC{G} = \chiC{U} = 1$ (cooperation with good and unknown recipients) and $\hast \to 1$ are required.
The logic is as follows: if $\hast < 1$, then achieving $p_c = 1$ would require $\chiC{B} = 1$ (cooperation with bad recipients), but this creates the Always Cooperate (ALLC) strategy, which is not evolutionarily stable.
When $\hast = 1$, Eq.~\eqref{eq:hast} simplifies to $\lp{1-q_f}\Rs{U} = 1 - q_f\Rs{G}$.
Since the left-hand side is bounded by $[0, 1-q_f]$ and the right-hand side is bounded by $[1-q_f, 1]$, equality requires $\Rs{U} = \Rs{G} = 1$.
Therefore, under cooperative ESS norms, donors must cooperate with both good and unknown recipients ($S(G) = S(U) = C$) and obtain good reputation with certainty ($R(G,C) = R(U,C) = 1$).

Since $S(B) \neq C$ is required to make the norm CESS, we must choose between $S(B) = D$ and $S(B) = P$.
The former yields \textit{CDC} norms (Cooperate-Defect-Cooperate for good, bad, and unknown recipients, respectively), and the latter yields \textit{CPC} norms (Cooperate-Punish-Cooperate for good, bad, and unknown recipients, respectively).

\subsubsection*{ESS condition for CDC norms}

To determine when CDC norms are evolutionarily stable, we derive the necessary and sufficient conditions by analyzing expected payoffs~\cite{murase2023indirect,murase2025costly}.

For a CDC norm to be ESS, cooperation must yield higher expected payoffs than alternative actions in each reputation context.
When facing a good recipient, the donor's expected payoff from cooperation must exceed that from defection and punishment.
Cooperation incurs immediate cost $c$ but generates future benefit $b\lb{q_f R\lp{G,C} + \lp{1-q_f}}$ when serving as a recipient, yielding a net payoff of $b\lb{q_f R\lp{G,C} + \lp{1-q_f}} - c$.
Defection and punishment yield $b\lb{q_f R\lp{G,D} + \lp{1-q_f}}$ and $b\lb{q_f R\lp{G,P} + \lp{1-q_f}} - \alpha$, respectively.
Cooperation dominates when:
\begin{equation}\label{eq:CDC_ESS_G}
    \begin{split}
        q_f b \lb{ R\lp{G,C} - R\lp{G,D} } &> c,\\
        q_f b \lb{ R\lp{G,C} - R\lp{G,P} } &> c - \alpha.
    \end{split}
\end{equation}
Analogous conditions apply for bad and unknown recipients:
\begin{equation}\label{eq:CDC_ESS_BU}
    \begin{split}
        q_f b \lb{ R\lp{B,D} - R\lp{B,C} } &> -c,\\
        q_f b \lb{ R\lp{B,D} - R\lp{B,P} } &> -\alpha, \\
        q_f b \lb{ R\lp{U,C} - R\lp{U,D} } &> c,\\
        q_f b \lb{ R\lp{U,C} - R\lp{U,P} } &> c - \alpha
    \end{split}
\end{equation}

The ESS conditions for CDC norms are thus determined by these six inequalities (Eqs.~\eqref{eq:CDC_ESS_G} and \eqref{eq:CDC_ESS_BU}).
For deterministic assessment rules ($R(X,A) \in \{0,1\}$), all viable CESS assessment rules are catalogued in Table~\ref{tab:CDC_ESS_norms}.
The table is organized by recipient reputation ($G$, $B$, $U$), and CESS norms are constructed by combining compatible assessment rules from each category.
For instance, by combining G1, B1, and U1 together with the CDC action rule, a CESS social norm that works when $q_f b > c$ is constructed.
There are $6 \times 4 = 24$ distinct cooperative ESS norms in total with the CDC action rule.

The analysis above shows that a CESS norm with the CDC action rule exists when
\begin{align}
  q_f b > c,
\end{align}
showing that lower $q_f$ requires a higher $b/c$ ratio to stabilize cooperation.
This result reproduces the analysis in~\cite{nakamura2011indirect}.

\begin{table}[!h]
\centering
\caption{\textbf{ESS combinations for the CDC norm.}
Entries show assessment rules $R(X,A)$ that can form CESS with $X\in\{G,B,U\}$ and $A\in\{C,D,P\}$.
The rightmost column lists the required parameter conditions.
By combining these rules, a CESS assessment rule is constructed.
For instance, an assessment rule G1-B6-U2 is CESS when $\alpha > q_f b > c$ and $q_f < 1$.
There are $24$ CESS norms with the CDC action rule in total.
}
\label{tab:CDC_ESS_norms}
\begin{tabular}{@{}l ccc l@{}}
\toprule
ID & $R(\cdot,C)$ & $R(\cdot,D)$ & $R(\cdot,P)$ & Conditions \\
\midrule
\multicolumn{5}{l}{\textit{G recipients}} \\
G1 & 1 & 0 & 0 & $q_f b>c$ \\
G2 & 1 & 0 & 1 & $q_f b>c$ and $\alpha>c$ \\
\hline
\addlinespace
\multicolumn{5}{l}{\textit{B recipients}} \\
B1 & 0 & 1 & 0 & any \\
B2 & 0 & 1 & 1 & any \\
B3 & 1 & 1 & 0 & any \\
B4 & 1 & 1 & 1 & any \\
B5 & 0 & 0 & 0 & $q_f < 1$ \\
B6 & 0 & 0 & 1 & $q_f < 1$ and $\alpha > q_f b$ \\
\hline
\addlinespace
\multicolumn{5}{l}{\textit{U recipients}} \\
U1 & 1 & 0 & 0 & $q_f b>c$ \\
U2 & 1 & 0 & 1 & $q_f b>c$ and $\alpha>c$ \\
\bottomrule
\end{tabular}
\end{table}

\subsubsection*{ESS condition for CPC norms}

We conduct a similar analysis for CPC norms.
Under CPC norms, $B$ recipients are punished by $\beta$ while $G$ recipients receive a benefit of $b$.
Thus, the difference between $G$ and $B$ recipients is $(b + \beta)$ rather than $b$.
ESS conditions are derived as follows:
\begin{equation}\label{eq:CPC_ESS}
    \begin{split}
        q_f (b+\beta) \lb{ R\lp{G,C} - R\lp{G,D} } &> c,\\
        q_f (b+\beta) \lb{ R\lp{G,C} - R\lp{G,P} } &> c - \alpha \\
        q_f (b+\beta) \lb{ R\lp{B,P} - R\lp{B,C} } &> \alpha-c,\\
        q_f (b+\beta) \lb{ R\lp{B,P} - R\lp{B,D} } &> \alpha, \\
        q_f (b+\beta) \lb{ R\lp{U,C} - R\lp{U,D} } &> c,\\
        q_f (b+\beta) \lb{ R\lp{U,C} - R\lp{U,P} } &> c - \alpha
    \end{split}
\end{equation}
This is analogous to Eqs.~\eqref{eq:CDC_ESS_G} and \eqref{eq:CDC_ESS_BU} for the CDC norms.

\begin{figure}[!h]
\centering
\includegraphics[width=0.6\textwidth]{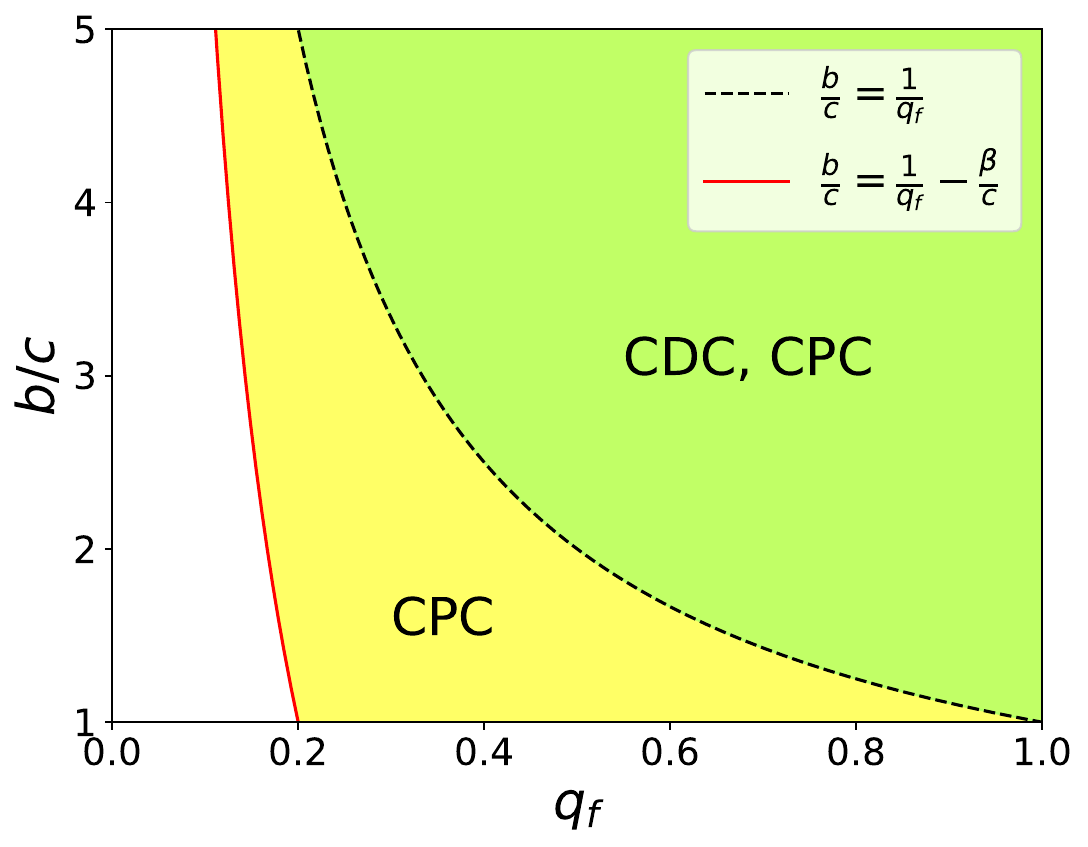}
\caption{
  The CESS parameter regions for the CDC and CPC norms.
  The green shaded area shows the parameter region where both CDC and CPC norms can maintain cooperation, $q_f b > c$.
  The yellow shaded area shows the parameter region where only CPC norms can maintain cooperation, $q_f (b+\beta) > \max\{c,\alpha\}$.
  The parameters used here are $\alpha / c = 0.5$ and $\beta / c = 4$.
}
\label{fig:ESS_CDC_CPC}
\end{figure}

When the norms are deterministic, the CPC norms that form ESS are summarized in Table~\ref{tab:CPC_ESS_norms}.
There exists at least one CESS norm when the following condition holds:
\begin{equation}
  q_f (b+\beta) > \max\{c,\alpha\}.
\end{equation}
Harsh punishment (higher $\beta$) and low punishment cost (lower $\alpha$) expand the parameter region where CESS norms exist.
The ESS parameter regions for CDC and CPC norms are compared in Fig.~\ref{fig:ESS_CDC_CPC}.

\begin{table}[!h]
\centering
\caption{\textbf{ESS combinations for the CPC norm.}
Entries show assessment rules $R(X,A)$ that can form CESS with CPC action rule $S(G) = C$, $S(B) = P$, and $S(U) = C$.
The rightmost column lists the required parameter conditions.
By combining these rules, a CESS assessment rule is constructed.
For instance, an assessment rule G1-B1-U1 is CESS when $q_f (b+\beta) > \max\{c,\alpha\}$.
Note that not all combinations are possible, unlike CDC norms.
B2 is incompatible with G2 and U2 as they have conflicting requirements: $\alpha > c$ and $c > \alpha$.
There are $5$ CESS norms in total.
}
\label{tab:CPC_ESS_norms}
\begin{tabular}{@{}l ccc l@{}}
\toprule
ID & $R(\cdot,C)$ & $R(\cdot,D)$ & $R(\cdot,P)$ & Conditions \\
\midrule
\multicolumn{5}{l}{\textit{G recipients}} \\
G1 & 1 & 0 & 0 & $q_f (b+\beta)>c$ \\
G2 & 1 & 0 & 1 & $q_f (b+\beta)>c$ and $\alpha>c$ \\
\hline
\addlinespace
\multicolumn{5}{l}{\textit{B recipients}} \\
B1 & 0 & 0 & 1 & $q_f (b+\beta) > \alpha$ \\
B2 & 1 & 0 & 1 & $q_f (b+\beta) > \alpha$ and $c > \alpha$ \\
\hline
\addlinespace
\multicolumn{5}{l}{\textit{U recipients}} \\
U1 & 1 & 0 & 0 & $q_f (b+\beta)>c$ \\
U2 & 1 & 0 & 1 & $q_f (b+\beta)>c$ and $\alpha>c$ \\
\bottomrule
\end{tabular}
\end{table}

\section*{Discussion}\label{sec:discussion}

In this study, we revisited the framework of indirect reciprocity under incomplete observation and reputation fading.
We showed that incomplete observation does not alter the conditions for cooperative evolution, as the reduced frequency of reputation updates is exactly offset by increased incentive strength.
In contrast, reputation fading creates qualitatively different and more challenging barriers to cooperation.
Specifically, we demonstrated that while classical norms like CDC (cooperate with good, defect against bad, cooperate with unknown) remain evolutionarily stable, CPC norms (where bad individuals are punished rather than simply defected against) expand the parameter region in which cooperation can be sustained.
This effect becomes particularly pronounced when punishment is harsh (high $\beta$) and the punishment cost is relatively small (low $\alpha$).
Importantly, including punishment allows broader ESS regions to exist at low benefit-to-cost ratios.
These CPC norms are as efficient as CDC norms since full cooperation is achieved.
As a general rule, individuals with uncertain reputations should be treated as if they are good, while those with definitively bad reputations should be punished, reminiscent of the ``innocent until proven guilty'' principle found in many legal systems.
We demonstrated that punishment is effective in mitigating the challenges posed by reputation fading when this principle is followed.

It may seem surprising and counterintuitive that incomplete observation does not affect the conditions for cooperation, as one might expect that less frequent reputation updates would make it harder to maintain cooperation.
However, our analysis shows that the increased incentive strength from longer-lasting reputations exactly compensates for less frequent updates.
When the observation probability $q_o$ is low, the expected duration of a reputation is longer, which increases the value of having a good reputation.
As long as reputations remain unchanged when actions are not observed, the conditions for cooperation remain unchanged.
This cancellation effect exists for third-order social norms as well.
We conjecture that this also holds for models with non-binary reputations, such as ternary~\cite{murase2022social,tanabe2013indirect} or continuous-valued reputations~\cite{lee2021local,lee2022second}, provided that reputations are shared without disagreement (i.e., under public assessment).
This is because the frequency of action observation only changes the effective time scale of the game, and we are interested in the long-term behavior at the stationary state.

We emphasize that this independence of the observation probability $q_o$ holds when players remain in the system for a long time compared to the time scale of reputation dynamics.
In our model, we assumed that players play indefinitely and that reputation dynamics reach a stationary state.
In reality, however, players may not stay in the system long enough.
If players do not play games for a sufficient duration compared to the time scale of reputation dynamics, they might not cooperate once assessed as $B$.
Thus, in reality, the observation probability $q_o$ should not be too small in such cases, so as to provide sufficient opportunities for $B$ players to be reassessed as $G$ and thereby incentivize cooperation.

We also emphasize that the independence of the observation probability $q_o$ does not hold for the private assessment model, where the opinions of players are not completely synchronized~\cite{murase2024indirect}.
In the private assessment model, the opinions are less synchronized as $q_o$ decreases, leading to instability of cooperation.
In particular, if $q_o \to 1/N$, the opinions may become completely independent.
It has been shown that no social norm can sustain evolutionarily stable cooperation in monomorphic populations under such conditions~\cite{murase2024indirect}.

\begin{figure}[!h]
\centering
\includegraphics[width=0.9\textwidth]{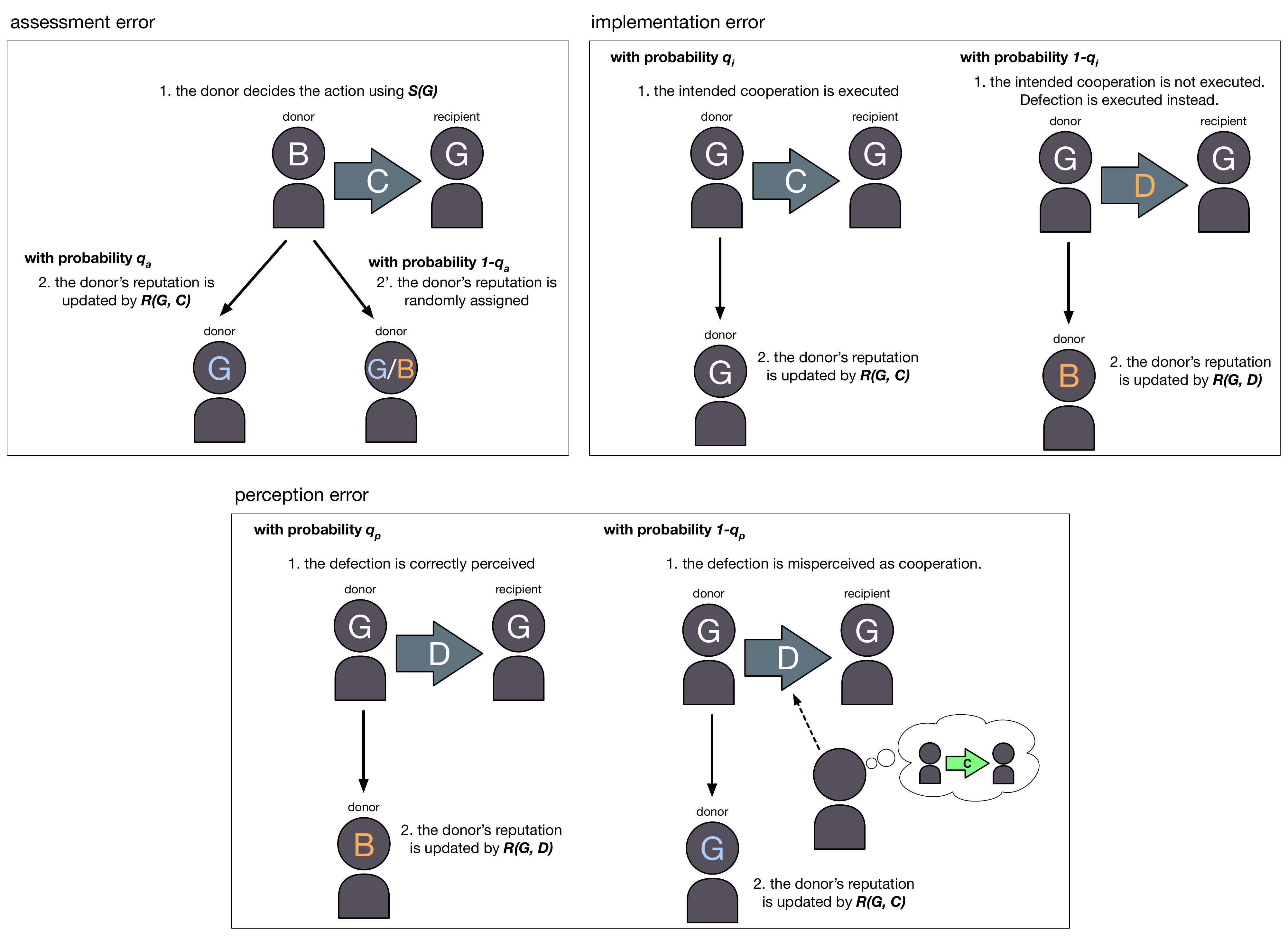}
\caption{
  Comparison of the three models with errors: assessment error, implementation error, and perception error.
  (Upper left) In the model with assessment error, the donor's reputation is not accurately updated.
  With probability $q_a$, the donor's reputation is correctly updated to $G$ or $B$; with probability $1-q_a$, the donor's reputation is randomly assigned to either $G$ or $B$.
  Here, $q_a \equiv 1 - 2 \mu$.
  (Upper right) In the model with implementation error, the donor's intended cooperation is not always carried out.
  With probability $q_i$, the intended cooperation is correctly carried out.
  With probability $1-q_i$, the donor may instead execute $D$.
  The accuracy parameter is defined as $q_i \equiv 1 - \mu_i$.
  (Bottom) In the model with perception error, the donor's defection is not accurately perceived.
  With probability $q_p$, the defection is correctly perceived and the reputation is updated according to $R(X, D)$.
  On the other hand, with probability $1-q_p$, the donor's defection is misperceived as cooperation and the reputation is updated according to $R(X, C)$.
  Here, $q_p \equiv 1 - \epsilon_{DC}$.
  While these differences may seem minor, they yield qualitatively different results.
}
\label{fig:three_models}
\end{figure}

While we consider two specific models of information limitations, other sources of inaccuracy merit consideration.
One of the most common types of noise is \textit{assessment error}, where a player's reputation is incorrectly assigned.
Another source of noise is \textit{implementation error}, where the donor's intended cooperation is not executed and defection occurs instead.
A third type is \textit{perception error}, where the action taken by the donor is not correctly perceived when the donor is assessed.
See Fig.~\ref{fig:three_models} for the definitions of these errors.
Generally, these errors make it harder to maintain cooperation by creating uncertainty in the reputation system.
For instance, for L3 and L6 social norms under the public assessment model, the required benefit-to-cost ratio~\cite{glynatsi2025exact} is
\begin{equation}
  \frac{b}{c} > \frac{1}{\lp{1 - 2\mu} \lp{1 - \mu_i} \lp{1 - \epsilon_{DC}}},
\end{equation}
where $\mu$ is the assessment error rate, $\mu_i$ is the implementation error rate, and $\epsilon_{DC}$ is the perception error rate with which the donor's defection is perceived as cooperation.
Punishment does not always help mitigate the impact of these errors.
For instance, when assessment error or implementation error is large, punishment can actually reduce overall welfare~\cite{ohtsuki2009indirect,murase2025costly}.
Since the reduction is so significant, the realized payoffs are often lower than those under ALLD norms, making punishment ineffective.
In contrast, punishment can be effective for reputation fading and perception error~\cite{murase2025costly}.

\begin{figure}[!h]
  \centering
  \includegraphics[width=0.9\textwidth]{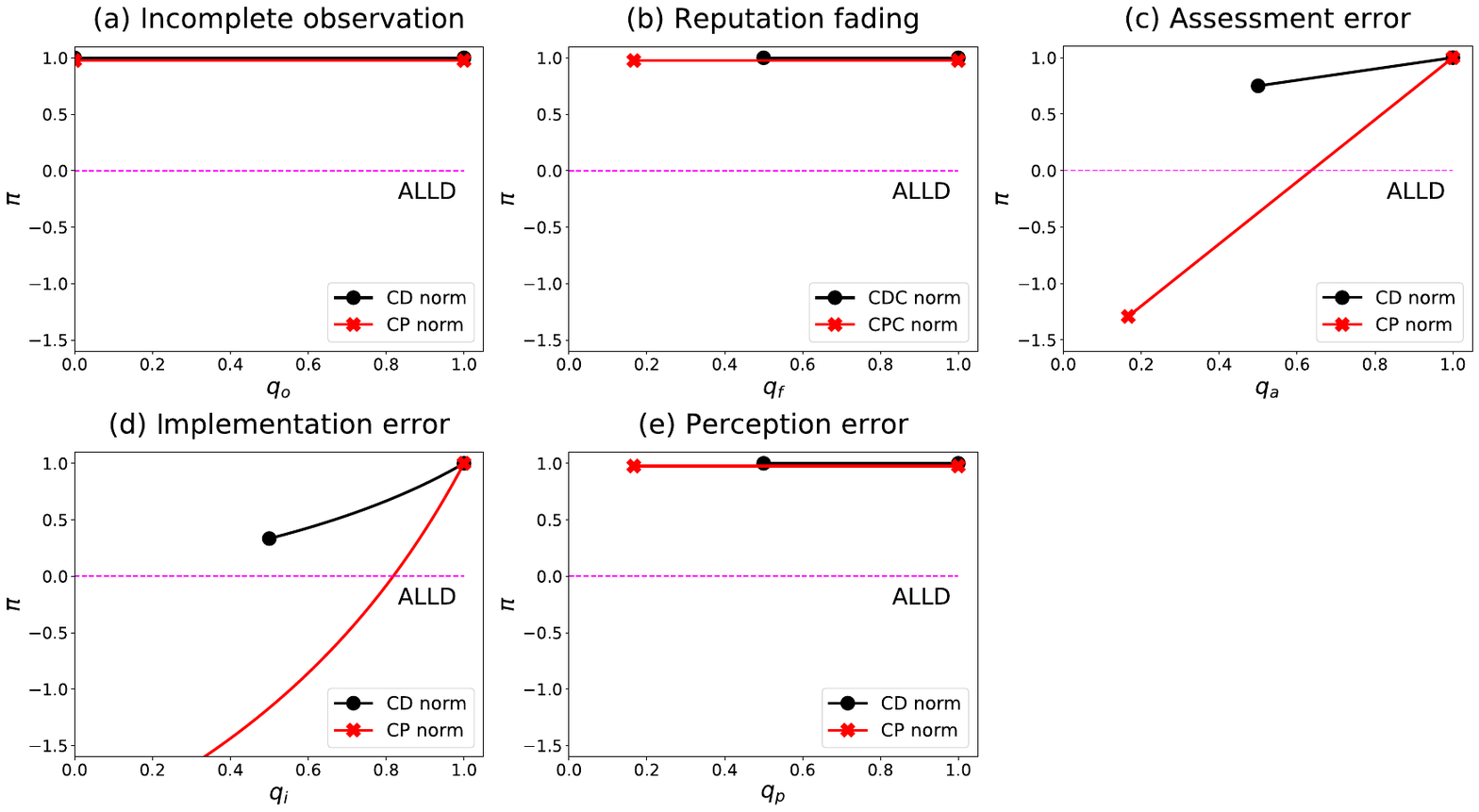}
  \caption{
    Comparison of the five models with different types of information inaccuracy: (a) incomplete observation, (b) reputation fading, (c) assessment error, (d) implementation error, and (e) perception error.
    The horizontal axis represents the accuracy of information ($q_o$, $q_f$, $q_a$, $q_i$, and $q_p$).
    The vertical axis represents the players' payoffs under the ESS social norms.
    The black and red solid curves represent the payoffs under the CD and CP norms, respectively.
    The curves are drawn only for $q$ values where the corresponding norm is ESS.
    The vertical dashed lines indicate $\pi = 0$, the payoff under the ALLD norm.
    For (c) and (d), the CP norm is ESS over a wider range of $q$, but the realized payoffs are lower than those under the ALLD norm.
    The parameters are $b$ = 2.0, $c$ = 1.0, $\alpha$ = 0.5, and $\beta$ = 4.0.
    }
  \label{fig:five_models_comparison}
\end{figure}

To highlight these different kinds of information inaccuracy, we compare the two models studied in this paper and the three types of noise studied in previous studies~\cite{ohtsuki2009indirect,murase2025costly,glynatsi2025exact}.
Since the latter three models were analyzed in the previous works, we only show the summary of the analysis in Appendix.
In Figure~\ref{fig:five_models_comparison}, we compare these five models to show the effectiveness and limitations of punishment under different types of information inaccuracy.
The horizontal axis represents the accuracy of information $q$ and the vertical axis represents the payoff of the players.
As the figures show, the introduction of punishment is effective for reputation fading and perception error.
It widens the parameter region where cooperation can be maintained without significant negative side effects.
On the other hand, for assessment and implementation errors, punishment is ineffective.
Even though the norm is evolutionarily stable for a wider range of $q$, the realized payoffs are often lower than those under ALLD norms, making punishment ineffective.
This is summarized in Table~\ref{tab:summary_noise}.

In general, punishment works well when inaccurate information does not reduce cooperation levels.
When inaccurate information leads to reduced cooperation, costly punishment can cause a significant decline in overall payoffs, making the introduction of punishment counterproductive~\cite{ohtsuki2009indirect}.
On the other hand, when inaccurate information does not lead to reduced cooperation, such as overlooking free-riders, costly punishment can be effective in mitigating the impact of inaccurate information without significant negative side effects.
In other words, while costly punishment can deter free-riders and mitigate the impact of inaccurate information, the punishment should not be executed often.

\begin{table}[!h]
\centering
\caption{
  Summary of the impact of different types of information inaccuracy on cooperation and the effectiveness of costly punishment.
  In the left column, we list the type of noise affecting the reputation system.
  In the middle column, we show the required benefit-to-cost ratio $b/c$ for cooperation to be sustained.
  In the right column, we indicate whether costly punishment is effective in mitigating the impact of the noise.
  Here, ``ineffective'' means that costly punishment leads to a significant decline in overall payoffs, while ``effective'' means that costly punishment can mitigate the impact of the noise without significant negative side effects.
  }
\label{tab:summary_noise}
\begin{tabular}{@{}l c c@{}}
\toprule
Type of noise & Required $b/c$ & Costly punishment \\
\midrule
\makecell[l]{Incomplete observation \\ \quad (Reputation remains unchanged)} & No change & Not necessary \\
\makecell[l]{Reputation fading \\ \quad (Reputation is set to $U$)} & Increase & Effective \\
\makecell[l]{Assessment error~\cite{ohtsuki2009indirect} \\ \quad (Reputation is randomized to $G$ or $B$)} & Increase & Ineffective \\
\makecell[l]{Implementation error~\cite{murase2025costly} \\ \quad (D is executed instead of C)} & Increase & Ineffective \\
\makecell[l]{Perception error~\cite{murase2025costly} \\ \quad (D is recognized as C)} & Increase & Effective \\
\bottomrule
\end{tabular}
\end{table}

For future work, it would be of particular interest to study private assessment models~\cite{hilbe2018indirect,murase2024indirect}.
In this paper, we assume that reputations are shared without disagreement.
Even when a reputation is unknown, both the donor and observers agree that the recipient's reputation is $U$.
However, this is a strong idealization in reality, especially when the population size is large.
To better understand the impact of reputation fading, it is also important to consider cases where reputations are not perfectly shared.
It would be particularly interesting to examine how introducing unknown reputation states affects cooperation in the private assessment model.

\bibliography{indirect}

\section*{Acknowledgement}

The authors would like to thank C.~Hilbe and S.~K.~Baek for insightful discussions and feedback on this work.
HK acknowledges support by Basic Science Research Program through the National Research Foundation of Korea (NRF) funded by the Ministry of Education (NRF-2020R1I1A2071670).
YM acknowledges support from JSPS KAKENHI Grant Number JP25K07145.
HK acknowledges support from the international internship program at RIKEN R-CCS, where part of this study was conducted.

\section*{Author contributions statement}

Y.M. conceived and designed the study. H.K. and Y.M. developed the model, performed the analysis, and wrote the manuscript.

\newpage

\section*{Appendix}

\subsection*{Formal analysis of third-order social norms under incomplete observation}\label{subsec:third_order_incomplete}

Suppose that a third-order assessment rule is represented by $R(X, Y, A)$, where $X \in \{ G, B \}$ is the donor's reputation, $Y \in \{ G, B \}$ is the recipient's reputation, and $A \in \{ C, D \}$ is the action taken by the donor.
A third-order action rule is represented by $S(X, Y)$, where $X \in \{ G, B \}$ is the donor's reputation and $Y \in \{ G, B \}$ is the recipient's reputation.

The previous analysis shows that a social norm is ESS if and only if certain conditions are satisfied~\cite{glynatsi2025exact}.
A third-order social norm with assessment rule $R(X,Y,A)$ and action rule $S(X, Y)$ is an ESS if and only if the following conditions hold:
\begin{equation}
  \label{eq:ESS}
  \begin{cases}
    \lb{R\lp{X,Y,C} - R\lp{X,Y,D}} \Delta v > c \quad &\text{if } S(X, Y) = C \\[0.2cm]
    \lb{R\lp{X,Y,C} - R\lp{X,Y,D}} \Delta v < c \quad &\text{if } S(X, Y) = D
  \end{cases}
\end{equation}
for all possible contexts $(X, Y) \in \big\{(G, G), (G, B), (B, G), (B, B)\big\}$.
Here, $\Delta v$, defined below, is the difference in the long-term expected payoffs between $G$ and $B$ players.
The left-hand side of the inequalities in Eq.~\eqref{eq:ESS} represents the expected long-term surplus between cooperation and defection associated with being assessed as $G$, while the right-hand side represents the instantaneous cost of cooperation.
Cooperation is favorable when the expected long-term surplus is larger than the cost of cooperation, while defection is favorable when the expected long-term surplus is smaller than the cost of cooperation.

The expected long-term payoff difference $\Delta v$ is defined as
\begin{equation}
  \label{eq:Delta_v}
  \Delta v = \frac{ b \lb{ \hast \chiC{G,\Delta} \!+\! \lp{1\!-\!\hast} \chiC{B,\Delta} }  -c \lb{ \hast \chiC{\Delta,G} \!+\! \lp{1\!-\!\hast} \chiC{\Delta,B} } }{ 1 - \hast \Rs{\Delta,G} - (1\!-\!\hast) \Rs{\Delta,B} }.
\end{equation}
Here, we use the following definitions for $X, Y \in \{G, B\}$
\begin{equation}
  \begin{split}
  \chiC{X, Y} &\equiv \begin{cases}
    1 &\text{if } S(X, Y) = C, \\
    0 &\text{if } S(X, Y) = D,
  \end{cases} \\
  \Rs{X, Y} &\equiv R(X, Y, S(X, Y)), \\
  \chiC{X, \Delta} &\equiv \chiC{X,G} - \chiC{X,B},  \\
  \chiC{\Delta, X} &\equiv \chiC{G,X} - \chiC{B,X},  \\
  \Rs{\Delta, X}   &\equiv \Rs{G, X} - \Rs{B, X}.
  \end{split}
\end{equation}
The equilibrium fraction of $G$ reputations in the population is denoted by $\hast$, which is calculated as the solution to the following quadratic equation
\begin{equation} \label{eq:quadratic}
  c_2 {\hast}^2 + c_1 \hast + c_0 = 0,
\end{equation}
where $c_2$, $c_1$, and $c_0$ are defined as
\begin{equation}\label{eq:c2c1c0}
  \begin{split}
    c_2 &\equiv \Rs{G,G} - \Rs{G,B} - \Rs{B,G} + \Rs{B,B}  \\
    c_1 &\equiv \Rs{G,B} + \Rs{B,G} -2\Rs{B,B} -1  \\
    c_0 &\equiv \Rs{B,B} \\
  \end{split}.
\end{equation}

When the action is assessed only with probability $q_o$, the effective assessment rule is rescaled as
\begin{equation}
    \label{eq:rescaling_q}
    R^{\dag}(X, Y, A) = q_o R(X, Y, A) + (1-q_o) \delta_{X, G},
\end{equation}
where $\delta_{X, G}$ is the Kronecker delta function that takes the value $1$ if $X = G$ and $0$ otherwise.
Hereafter, we denote variables for the incomplete observation case ($q_o < 1$) with a dagger.

The rescaling in Eq.~\eqref{eq:rescaling_q} does not alter the solution of the quadratic equation in Eq.~\eqref{eq:quadratic}.
This is because the coefficients in Eq.~\eqref{eq:c2c1c0} are rescaled as
\begin{equation}
  \begin{split}
    c_2^{\dagger} &= q_o \lb{\Rs{G,G} - \Rs{G,B} - \Rs{B,G} + \Rs{B,B}}  \\
    c_1^{\dagger} &= q_o \lb{\Rs{G,B} + \Rs{B,G} -2\Rs{B,B} -1 }  \\
    c_0^{\dagger} &= q_o \lb{\Rs{B,B} } \\
  \end{split}.
\end{equation}
Thus, the solution remains the same:
\begin{equation}
  h^{\ast \dagger} = \hast
\end{equation}

Next, consider the rescaling of Eq.~\eqref{eq:Delta_v}.
The numerator remains unchanged by the rescaling since $\hast$ and the action rule remain the same.
The denominator is rescaled by $q_o$ because
\begin{equation}
    \begin{split}
    \Rs{\Delta, G}^{\dagger} = R_S^{\dagger}\lp{G, G} - R_S^{\dagger}\lp{G, G} = q_o \Rs{\Delta, G} + (1-q_o) \\
    \Rs{\Delta, B}^{\dagger} = R_S^{\dagger}\lp{G, B} - R_S^{\dagger}\lp{G, B} = q_o \Rs{\Delta, B} + (1-q_o).
    \end{split}
\end{equation}
Therefore, the marginal benefit of being good is
\begin{equation}
  \Delta v^{\dagger} = \Delta v / q_o.
\end{equation}

As a result, the ESS conditions in Eq.~\eqref{eq:ESS} are also unchanged.
Plugging the rescaled assessment rule $R^{\dag}(X, Y, A)$ and $\Delta v^{\dagger}$ into Eq.~\eqref{eq:ESS} yields
\begin{equation}
    \begin{split}
    \lb{R^{\dag}\lp{X,Y,C} - R^{\dag}\lp{X,Y,D}} \Delta v^{\dag} &= q_o \lb{R\lp{X,Y,C} - R\lp{X,Y,D}} \Delta v / q_o + (1-q_o) \lb{\delta_{X, G} - \delta_{X,G}} \Delta v / q_o \\
    &= \lb{R\lp{X,Y,C} - R\lp{X,Y,D}} \Delta v.
    \end{split}
\end{equation}
Thus, the ESS condition and the cooperation level are independent of $q_o$.

When the action is observed with probability $q_o < 1$, the reputations are updated less frequently.
This increases the incentive $\Delta v$ to be assessed as $G$ by a factor of $1/q_o$ relative to perfect observation.
This is because it takes $1/q_o$ times longer to recover a $G$ reputation once a player is assessed as $B$.
On the other hand, the probability of being assessed as $B$ becomes $q_o$ times smaller.
Those two effects cancel each other out, which makes the ESS conditions and the cooperation level independent of $q_o$.

More generally, $q_o$ only changes the time scale of the game and does not change the ESS conditions or the cooperation level because we consider indefinitely repeated games.

\subsection*{Assessment error, implementation error, and perception error}\label{subsec:errors}

Here, we consider models with non-vanishing error rates.
More specifically, we consider three types of errors: assessment error, implementation error, and perception error.
Since these were studied in previous work~\cite{ohtsuki2009indirect,murase2025costly,glynatsi2025exact}, we only summarize the results here.

In the presence of these errors, the effective benefit, cost, and assessment rules are rescaled.
The introduction of these errors is equivalent to rescaling $b \to \tilde{b}$, $c \to \tilde{c}$, and $R(X,A) \to \tilde{R}(X,A)$.
First, we show the results for the general case, leaving the concrete forms of the rescaling for later.

The ESS conditions for CD-norms can be written as
\begin{equation}\label{eq:CD_ESS}
  \begin{split}
    \tilde{b} \lb{ \Rtilde{G,C} - \Rtilde{G,D} } &> \tilde{c},\\
    \tilde{b} \lb{ \Rtilde{G,C} - \Rtilde{G,P} } &> \tilde{c} - \alpha, \\
    \tilde{b} \lb{ \Rtilde{B,D} - \Rtilde{B,C} } &> -\tilde{c},  \\
    \tilde{b} \lb{ \Rtilde{B,D} - \Rtilde{B,P} } &> -\alpha.
  \end{split}
\end{equation}
Similarly, the ESS conditions for CP-norms are
\begin{equation}\label{eq:CP_ESS}
  \begin{split}
    \lp{\tilde{b} + \beta} \lb{ \Rtilde{G,C} - \Rtilde{G,D} } &> \tilde{c},\\
    \lp{\tilde{b} + \beta} \lb{ \Rtilde{G,C} - \Rtilde{G,P} } &> \tilde{c} - \alpha, \\
    \lp{\tilde{b} + \beta} \lb{ \Rtilde{B,P} - \Rtilde{B,C} } &> \alpha -\tilde{c},  \\
    \lp{\tilde{b} + \beta} \lb{ \Rtilde{B,P} - \Rtilde{B,D} } &> \alpha.
  \end{split}
\end{equation}

The average fraction of $G$-players in the population is
\begin{equation}
  \hast = \frac{ \Rtilde{B, S\lp{B}} }{ 1 - \Rtilde{G, S\lp{G}} + \Rtilde{B, S\lp{B}} }.
\end{equation}
The payoffs for the CD-norms and CP-norms are
\begin{equation}\label{eq:payoff_CD_CP}
  \begin{split}
    \pi_{CD} &= \hast \lp{\tilde{b}-\tilde{c}},  \\
    \pi_{CP} &= \hast \lp{\tilde{b}-\tilde{c}} + \lp{1-\hast} \lp{-\alpha - \beta}.
  \end{split}
\end{equation}

As we will see later, one of the norms with the highest cooperation level and widest ESS parameter region is the analogue of the Stern-Judging norm.
The definitions of the SJ-like CD-norm and CP-norm are summarized in Table~\ref{tab:SJ_norms}.

\begin{table}[!h]
\centering
\caption{
  The definitions of the SJ-like CD-norm and CP-norm.
  Those norms have the highest cooperation level and the widest ESS parameter region among CD-norms and CP-norms, respectively, for the three models with non-vanishing errors.
}
\label{tab:SJ_norms}
\begin{tabular}{@{}l cccccc @{}}
\toprule
Action rule & $R(G,C)$ & $R(G,D)$ & $R(G,P)$ & $R(B,C)$ & $R(B,D)$ & $R(B,P)$ \\
\midrule
CD & 1 & 0 & 0 & 0 & 1 & 0 \\
CP & 1 & 0 & 0 & 0 & 0 & 1 \\
\bottomrule
\end{tabular}
\end{table}

\subsubsection*{Assessment errors}

Next, we consider more specific models.
First, we consider a model with non-vanishing assessment error rate $\mu > 0$ (but without implementation or perception errors).
We introduce the social resolution $q_a \equiv 1 - 2\mu$, which represents the probability that the reputation is correctly assigned.
Otherwise, with probability $1-q_a$, the reputation is randomly assigned to either $G$ or $B$.

Because of the assessment error, the effective assessment rule is rescaled as
\begin{equation}
  \begin{split}
  \tilde{R}(X,A) &\equiv \lp{1 - 2\mu} R(X,A) + \mu,  \\
                &= q_a R(X,A) + \lp{1-q_a}/2,
  \end{split}
\end{equation}
for $X \in \{G,B\}$ and $A \in \{C,D,P\}$.
The effective benefit and cost remain the same: $\tilde{b} = b$ and $\tilde{c} = c$.

By substituting the rescaled assessment rule into Eqs.~\eqref{eq:CD_ESS} and \eqref{eq:CP_ESS}, we obtain the ESS conditions for the CD-norms and CP-norms under assessment errors.
According to this analysis, ESS norms exist when
\begin{equation}
  \begin{cases}
    q_a b > c & \text{for CD-norms} \\
    q_a \lp{b+\beta} > \max\{c, \alpha\} & \text{for CP-norms}
  \end{cases}
\end{equation}
Among those ESS norms, the most efficient ones have $\hast = \lp{1-\mu} = \lp{1+q_a}/2$.
The highest payoffs are
\begin{equation}
  \begin{split}
    \pi_{CD} &= \lp{1+q_a} \lp{b-c} / 2,  \\
    \pi_{CP} &= \lp{1+q_a} \lp{b-c} / 2 + \lp{1-q_a} \lp{-\alpha - \beta} / 2, \\
             &= \lp{b-c-\alpha-\beta} / 2 + q_a \lp{b-c+\alpha+\beta} / 2.
  \end{split}
\end{equation}

\subsubsection*{Implementation error}\label{subsec:implementation_error}

Here, we consider a model where intended cooperation is not always executed.
With probability $(1-\mu_i)$, intended cooperation is successfully executed, while with probability $\mu_i$, defection occurs instead.
We define $q_i \equiv 1 - \mu_i$ as the probability of successful execution of the intended cooperation.
To isolate the effects of other errors, we consider the vanishing assessment error $\mu \to 0$.

The effect of the implementation error is taken into account by rescaling the effective assessment rule, benefit, and cost:
\begin{equation}
\begin{split}
  \Rtilde{X, C} &= (1 - \mu_i) R(X, C) + \mu_i R(X, D), \\
  \Rtilde{X, D} &= R(X, D),  \\
  \Rtilde{X, P} &= R(X, P),  \\
  \tilde{b} &= (1 - \mu_i) b, \\
  \tilde{c} &= (1 - \mu_i) c.
\end{split}
\end{equation}

By substituting the rescaled assessment rule into Eqs.~\eqref{eq:CD_ESS} and \eqref{eq:CP_ESS}, we obtain the ESS conditions for the CD-norms and CP-norms under non-vanishing implementation errors.
According to this analysis, ESS norms exist when
\begin{equation}
  \begin{cases}
    q_i b > c & \text{for CD-norms} \\
    q_i b+\beta > \max\{c, \alpha\} & \text{for CP-norms}
  \end{cases}
\end{equation}
Among those ESS norms, the most efficient ones have $\hast = 1 / \lp{2-q_i}$.
The highest payoffs are
\begin{equation}
  \begin{split}
    \pi_{CD} &= \frac{q_i \lp{b-c} }{2 - q_i},  \\
    \pi_{CP} &= \frac{q_i \lp{b-c} }{2 - q_i} + \frac{1-q_i}{2-q_i} \lp{-\alpha - \beta}.
  \end{split}
\end{equation}

\subsubsection*{Perception error}\label{subsec:perception_error}

Lastly, we consider a perception error model where the donor's defection is not correctly identified.
With probability $1-\epsilon_{DC}$, the donor's defection is correctly identified as defection, while with probability $\epsilon_{DC}$, the donor's defection is perceived as cooperation.
Let us define $q_p \equiv 1 - \epsilon_{DC}$ as the probability of correct perception of defection.

The effect of the perception error is taken into account by rescaling the effective assessment rule, benefit, and cost:
\begin{equation}
\begin{split}
  \Rtilde{X, C} &= R(X, C),  \\
  \Rtilde{X, D} &= (1-q_p) R(X, C) + q_p R(X, D),  \\
  \Rtilde{X, P} &= R(X, P),  \\
  \tilde{b} &= b, \\
  \tilde{c} &= c.
\end{split}
\end{equation}

By substituting the rescaled assessment rule into Eqs.~\eqref{eq:CD_ESS} and \eqref{eq:CP_ESS}, we obtain the ESS conditions for the CD-norms and CP-norms under non-vanishing perception errors.
According to this analysis, ESS norms exist when
\begin{equation}
  \begin{cases}
    q_p b > c & \text{for CD-norms} \\
    q_p \lp{b+\beta} > \max\{c, \alpha\} & \text{for CP-norms}
  \end{cases}
\end{equation}
Among those ESS norms, the most efficient ones have $\hast = 1$.
The highest payoffs are
\begin{equation}
  \begin{split}
    \pi_{CD} &= b-c,  \\
    \pi_{CP} &= b-c.
  \end{split}
\end{equation}

\nolinenumbers
\end{document}